\documentclass[journal]{IEEEtran}
\usepackage{amsmath,amssymb}
\usepackage{subfigure}
\usepackage{graphicx,graphics,color,psfrag}
\usepackage{cite,balance}
\usepackage{caption}
\allowdisplaybreaks
\usepackage{algorithm}
\usepackage{algorithmic}
\usepackage{accents}
\usepackage{amsthm}
\usepackage{bm}
\usepackage{url}
\usepackage[english]{babel}
\usepackage{multirow}
\usepackage{enumerate}
\usepackage{cases}
\usepackage{stfloats}
\usepackage{dsfont}
\usepackage{color,soul}
\usepackage{amsfonts}
\usepackage{cite,graphicx,amsmath,amssymb}
\usepackage{subfigure}
\usepackage{fancyhdr}
\usepackage{hhline}
\usepackage{graphicx,graphics}
\usepackage{array,color}
\usepackage{mathtools}
\usepackage{amsmath}

\newtheorem{example}{\bf Example}

\def\l{\left}
\def\r{\right}
\def\({\left(}
\def\){\right)}

\def\b0{{\mathbf{0}}}

\renewcommand{\mod}{\tx{mod}}

\newcommand{\tx}[1]{\texttt{#1}}

\newcommand{\diag}{\mathrm{diag}}

\newcommand{\nn}{\nonumber}

\begin{document}
\captionsetup[figure]{name={Fig.}}

\title{\huge 
Fast Beam  Training for IRS-Assisted \\ Multiuser Communications} \vspace{-3pt}
\!\!\!\author{Changsheng You,~\emph{Member,~IEEE}, Beixiong Zheng,~\emph{Member,~IEEE},
	 and Rui Zhang,~\emph{Fellow,~IEEE} 
	 \vspace{-20pt}
	   \thanks{\noindent The authors are with the Department of Electrical and Computer Engineering, National University of Singapore 117583, Singapore (Email: \{eleyouc, elezbe, elezhang\}@nus.edu.sg).
	\vspace{-12pt}
}}  
\maketitle

\begin{abstract}
In this letter, we consider an intelligent reflecting surface (IRS)-assisted multiuser communication system, where an IRS is deployed to provide virtual line-of-sight (LoS) links between an access point (AP) and multiple users. We consider the practical codebook-based IRS passive beamforming and study efficient design for IRS reflect beam training, which is challenging due to the large number of IRS reflecting elements. In contrast to the conventional single-beam training,   
we propose a new {\it multi-beam training} method by dividing the  IRS reflecting elements into multiple sub-arrays and designing their simultaneous  multi-beam steering over time. By simply comparing the received signal power over time, 
each user can detect its optimal IRS beam direction with a high probability, even without searching over all possible beam directions as the single-beam training. Simulation results show that  our proposed multi-beam training  significantly reduces the training time of conventional
 single-beam training and yet achieves comparable IRS passive beamforming performance for data transmission.

\end{abstract}
\vspace{-5pt}
\vspace{-11pt}
\section{Introduction}
Intelligent reflecting surface (IRS) has emerged as a promising cost-effective technology for enhancing the spectral and energy efficiency of future wireless networks \cite{qingqing2019towards}. In particular, by smartly controlling signal reflection  via a massive number of low-cost passive reflecting elements, IRS is able to dynamically program the radio propagation environment for achieving signal enhancement and/or interference suppression. Compared to traditional active relay, IRS incurs much lower hardware cost and energy consumption due to its passive reflection. 
These appealing advantages have spurred intensive enthusiasm  recently in deploying IRS to enhance the communication performance of various  wireless systems   \cite{zheng2020intelligent,you2019progressive,zhaoMM2019,ning2019channel,tan2018enabling}.

Particularly, for millimeter-wave (mmWave) communications at high operation frequencies where the direct channels between an access point (AP) and its served  users are susceptible to severe blockage and propagation loss,  IRS can be properly deployed to provide virtual line-of-sight (LoS) AP-IRS-user links, and hence significantly enhance their communication performance  \cite{tan2018enabling}. To reap the large passive beamforming gain of IRS, it is indispensable for 
the IRS to conduct passive/reflect  beam training in coordination with the AP's transmit beam training for establishing high signal-to-noise ratio (SNR) links with IRS-assisted users before implementing efficient  channel estimation and data transmission. This, however,  is practically challenging due to the massive number of IRS reflecting elements that generate pencil-like sharp beams and thus require a large number of beam directions in the training codebook to cover the space of interest. The conventional single-beam training needs to search over all possible beam directions and inevitably  incurs  prohibitively high training overhead.

\begin{figure}[t]
\vspace{5pt}
\begin{center}
\includegraphics[height=3.4cm]{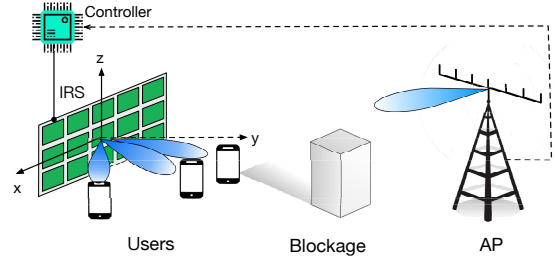}
\caption{IRS-assisted multiuser communication system.}
\label{Fig:Syst}
\end{center}
\vspace{-5pt}
\end{figure}
This thus motivates this letter to study a more efficient beam training design for  an IRS-assisted multiuser communication system as shown in Fig.~\ref{Fig:Syst}, where an IRS is deployed to establish LoS   links with both a multi-antenna AP and a group of  single-antenna users that are assumed to be located near the helping IRS (e.g., in a hot spot scenario).  
As the AP and IRS are at fixed locations, we assume for simplicity that the AP's transmit beamforming is fixed and thereby focus  
on designing the reflect  beam training for IRS. To reduce the training time of conventional single-beam training, we propose a new {\it multi-beam training} method. Specifically, we divide the  IRS reflecting elements into multiple sub-arrays and design their multi-beam codebook to steer different beam directions simultaneously over time.
Then, each user can detect its optimal IRS beam direction with a high probability via simple received  signal power/SNR comparisons over time, without the need of searching over all possible beam directions as the single-beam training.
 Simulation results show significant training time  reduction by our proposed multi-beam training as compared to conventional  single-beam training, yet  without compromising much the  IRS passive beamforming performance for data transmission. It is worth noting that multi-beam training design has also been proposed in \cite{hassanieh2018fast}; however, it relies on random hashing (RH), which incurs a random number of training symbols for achieving unique beam identification. In contrast, our proposed multi-beam training applies with any given number of training symbols and is numerically shown to outperform that in \cite{hassanieh2018fast} in terms of passive beamforming gain given the same training time.

\vspace{-8pt}
\section{System Model}\label{Sec:Model}
\vspace{-3pt}

As shown in  Fig.~\ref{Fig:Syst}, we consider the downlink beam training in an IRS-assisted  multiuser  communication system,
  where an IRS  is deployed to assist the communications between an AP equipped with an $N_{\rm A}$-antenna  uniform linear array (ULA)   and $K$ single-antenna users, denoted by the set $\mathcal{K}=\{1, 2,\cdots, \!K\}$.  The IRS is composed of $N_{\rm I}\!=\!N_{\rm x}\!\times\! N_{\rm z}$ reflecting elements placed in the $x$-$z$ plane,  and is attached with a smart controller for  tuning signal reflection at each reflecting element as well as exchanging information with the AP via a separate reliable link. The users are distributed in the same horizontal $x$-$y$ plane with the IRS located  in the center.

We consider the propagation environment with  limited scattering (which is typical for mmWave channels) and  adopt the commonly-used geometric channel model \cite{el2014spatially}. Assume that the direct AP-user links are blocked due to obstacles (e.g., buildings), whereas by properly deploying the IRS,  there exists a deterministic LoS  path in both the AP-IRS and IRS-user links. Let ${\bf u}(\phi, N)$ denote the \emph{steering vector function}, defined as
 \vspace{-12pt}
\begin{align}
{\bf u}(\phi, N)&\triangleq\l[1, e^{-\jmath \pi 1 \phi},\cdots, e^{-\jmath\pi (N-1) \phi}\r]^{T},\label{Eq:SteerVec}
\vspace{-10pt}
\end{align}
 where $N$ is the ULA array size, $\phi$ denotes the constant phase difference between the observations at two adjacent antennas/elements, and $\jmath$ denotes the imaginary unit. Since the AP and IRS are at fixed locations once deployed, the AP-IRS link typically has a much longer channel coherence time than the IRS-user link (due to user mobility) and thus can be considered as quasi-static.
As such, we assume for simplicity that the AP has aligned its transmit beamforming with the AP-IRS LoS channel and thus can be treated as having an equivalent single antenna.
 Then, the effective channel from the AP to IRS, denoted by ${\bf h}\in {\mathbb C}^{N_{\rm I}\times 1}$, can be modeled as ${\bf h}=h {\bf a}_{\rm r}(\theta_{\rm I}^{\rm r},\vartheta_{\rm I}^{\rm r})$.
where $h$ denotes the complex-valued path gain of the AP-IRS link; $\theta_{\rm I}^{\rm r}\in[0, \pi]$ and $\vartheta_{\rm I}^{\rm r}\in[0, \pi]$ denote respectively the (physical) azimuth and elevation angles-of-arrival (AoAs) at the IRS.
Moreover, ${\bf a}_{\rm r}\in\mathbb{C}^{N_{\rm I}\times 1}$ represents the  receive array response vector of IRS, which can be expressed as
${\bf a}_{\rm r}(\theta_{\rm I}^{\rm r}, \vartheta_{\rm I}^{\rm r})={\bf u}(\phi_{\rm I}^{\rm r}, N_{\rm x})\otimes {\bf u}(\psi_{\rm I}^{\rm r}, N_{\rm z})$,
where $\otimes$ stands for the Kronecker product, $\phi_{\rm I}^{\rm r}\triangleq\frac{2 d_{\rm I}}{\lambda}\cos(\theta_{\rm I}^{\rm r})\sin(\vartheta_{\rm I}^{\rm r})\in [-\frac{2 d_{\rm I}}{\lambda}, \frac{2 d_{\rm I}}{\lambda}]$ and
 $\psi_{\rm I}^{\rm r}\triangleq\frac{2 d_{\rm I}}{\lambda}\cos(\vartheta_{\rm I}^{\rm r})\in [-\frac{2 d_{\rm I}}{\lambda}, \frac{2 d_{\rm I}}{\lambda}]$ are referred to as the horizontal and vertical \emph{spatial directions}, respectively,
 with $\lambda$   and $d_{\rm I}$ respectively denoting the signal wavelength and IRS's  reflecting element spacing. Note that there exists a one-to-one mapping between $\{\phi_{\rm I}^{\rm r}, \psi_{\rm I}^{\rm r}\}$ and $\{\theta_{\rm I}^{\rm r}, \vartheta_{\rm I}^{\rm r}\}$. 
Moreover, due to the same altitude of the IRS center and users,  the elevation angles-of-departure (AoD) from  the IRS to different users are all $\pi/2$ and denoted by $\vartheta_{{\rm I},k}^{\rm t}\!\triangleq\vartheta_{{\rm I}}^{\rm t}\!=\! \pi/2, \!\forall k\in\mathcal{K}$.
 Then, the IRS-user~$k$ LoS path can be modeled as ${\bf g}^{H}_k=g_k  {\bf b}_{\rm t}^{H}(\theta_{{\rm I},k}^{\rm t},\vartheta_{{\rm I}}^{\rm t}), \forall k\in\mathcal{K}$, where $g_k$ denotes the complex-valued path gain of the IRS-user $k$ link;  $\theta_{{\rm I},k}^{\rm t}\in[0,\pi]$ denotes the azimuth AoD from  the IRS to user $k$;  and ${\bf b}_{\rm t}(\theta_{{\rm I},k}^{\rm t},\vartheta_{{\rm I}}^{\rm t})={\bf u}(\phi_{{\rm I},k}^{\rm t}, N_{\rm x})\otimes {\bf u}(\psi_{{\rm I}}^{\rm t}, N_{\rm z})$ represents the transmit array response vector of IRS with $\phi_{{\rm I},k}^{\rm t}=\frac{2  d_{\rm I}}{\lambda}\cos(\theta_{{\rm I},k}^{\rm t})\sin(\vartheta_{{\rm I},k}^{\rm t})$ and $\psi_{{\rm I}}^{\rm t}=\frac{2  d_{\rm I}}{\lambda}\cos(\vartheta_{{\rm I}}^{\rm t})$. 

Let $\boldsymbol{\Omega}\triangleq\diag(e^{\jmath\omega_1}, e^{\jmath\omega_2},\cdots, e^{\jmath\omega_{N_{\rm I}}})\in\mathbb{C}^{N_{\rm I}\times N_{\rm I}}$ denote the  diagonal IRS reflecting matrix, where for simplicity we assume that the reflection amplitude of each element is set to one (or its maximum value),{\footnote{This assumption is valid for the ideal case of independent  reflection amplitude-and-phase control; while for the practical case of phase-dependent amplitude control \cite{abeywickrama2020intelligent}, we need to assume that the effective resistance of each reflecting element is sufficiently low so that its reflection amplitude variation over phase is negligible.}} and $\omega_n$, $n\in\{1,2,\cdots, N_{\rm I}\}$  denotes the reflection phase shift of element $n$.\footnote{The proposed IRS beam-training method can be extended to the case with practical IRS discrete phase shifts by e.g., using the nearest-phase quantization for discretizing the continuous phase shifts as in \cite{you2019progressive}.}
Based on \cite{you2019progressive}, the received signal at each user $k$ is given by
\begin{align}
\vspace{-7pt}
\!\!y_k&= {\bf g}_{k}^{H}\boldsymbol{\Omega}{\bf h}x+{n}_k\nn \\
\vspace{-2pt}
&=h g_k {\bf b}_{\rm t}^{H}(\theta_{{\rm I},k}^{\rm t},\vartheta_{{\rm I}}^{\rm t})\boldsymbol{\Omega} {\bf a}_{\rm r}(\theta_{\rm I}^{\rm r}, \vartheta_{\rm I}^{\rm r})x+{n}_k \nn\\
\vspace{-2pt}
&=\eta_k {\bf c}^{H}_{k} {\bf v} x+n_k, \qquad\qquad\forall k\in\mathcal{K},\label{Eq:Signal}
\vspace{-6pt}
\end{align}
where  $x\in \mathbb{C}$ denotes the symbol transmitted by the AP with power $P_{\rm A}$, ${n}_k$ is the received additive white Gaussian noise (AWGN) at user $k$ with power $\sigma^2$, $\eta_k\triangleq h g_k$, ${\bf v}\triangleq[e^{j\omega_1}, e^{j\omega_2},\cdots, e^{j\omega_{N_{\rm I}}}]^{T}$, and 
\vspace{-3pt}
\begin{align}
{\bf c}^{H}_{k}&\triangleq {\bf b}_{\rm t}^{H}(\theta_{{\rm I},k}^{\rm t},\vartheta_{{\rm I}}^{\rm t})\odot  {\bf a}^{T}_{\rm r}(\theta_{\rm I}^{\rm r}, \vartheta_{\rm I}^{\rm r})\nn\\
&=\!({\bf u}^{H}\!(\phi_{{\rm I},k}^{\rm t}, N_{\rm x})\!\otimes\! {\bf u}^{H}\!(\psi_{{\rm I}}^{\rm t}, N_{\rm z}))\!\odot\! ({\bf u}^{T}\!(\phi_{\rm I}^{\rm r}, N_{\rm x})\!\otimes\! {\bf u}^{T}\!(\psi_{\rm I}^{\rm r}, N_{\rm z}))\nn\\
&=\!({\bf u}^{H}\!(\phi_{{\rm I},k}^{\rm t}, N_{\rm x})\!\odot\!{\bf u}^{T}\!(\phi_{\rm I}^{\rm r}, N_{\rm x}))\!\otimes\! ({\bf u}^{H}\!(\psi_{{\rm I}}^{\rm t}, N_{\rm z}) \!\odot\! {\bf u}^{T}\!(\psi_{\rm I}^{\rm r}, N_{\rm z}))\nn\\
&\triangleq{\bf u}^{H}(\tilde{\varphi}_{{\rm I},k}, N_{\rm x}) \otimes  {\bf u}^{H}(\tilde{\chi}_{{\rm I}}, N_{\rm z}),
\end{align}
where  $\odot$ stands for the Hadamard product; $\tilde{\varphi}_{{\rm I},k}\triangleq\phi_{{\rm I},k}^{\rm t}-\phi_{\rm I}^{\rm r}\in[-\frac{4  d_{\rm I}}{\lambda}, \frac{4  d_{\rm I}}{\lambda}], \forall k\in\mathcal{K}$; and $\tilde{\chi}_{{\rm I}}\triangleq\psi_{{\rm I}}^{\rm t}-\psi_{\rm I}^{\rm r}\in[-\frac{4  d_{\rm I}}{\lambda}, \frac{4  d_{\rm I}}{\lambda}]$. 
Then, by leveraging the property that ${\bf u}(\phi, N)$ is a periodic function with period $2 $, 
 we define $\varphi_{{\rm I},k}\triangleq\tilde{\varphi}_{{\rm I},k} (\mod ~2)\in[-1, 1], \forall k\in\mathcal{K}$ as the \emph{effective cascaded}  IRS azimuth spatial direction  for each user $k$,  and $\chi_{{\rm I}}\triangleq\tilde{\chi}_{{\rm I}} (\mod ~2)\in[-1, 1]$
as  the \emph{common} IRS elevation spatial direction  for all the  users, such that 
 ${\bf u}(\varphi_{{\rm I},k}, N_{\rm x})={\bf u}(\tilde{\varphi}_{{\rm I},k}, N_{\rm x}), \forall k\in\mathcal{K}$, and ${\bf u}(\chi_{{\rm I}}, N_{\rm z})={\bf u}(\tilde{\chi}_{{\rm I}}, N_{\rm z})$,  where $a_1 (\mod~ a_2)$ denotes the modulo operation that returns the remainder after the division of $a_1$\! by $a_2$.
 
For the IRS reflect beam training, it can be easily observed from \eqref{Eq:Signal} that for each user $k$, the optimal  IRS beamforming vector is ${\bf v}={\bf c}_k={\bf u}({\varphi}_{{\rm I},k}, N_{\rm x}) \otimes  {\bf u}({\chi}_{{\rm I}}, N_{\rm z})$, i.e., both the azimuth and elevation directions are perfectly aligned. To reduce the computational complexity for the joint three-dimensional (3D) IRS beam training, we first write ${\bf v}$ as a Hadamard product of two vectors, i.e., ${\bf v}={\bf v}_{\rm x}\otimes {\bf v}_{\rm z}$, where ${\bf v}_{\rm x}=[e^{\jmath\omega_1}, e^{\jmath\omega_2}, \cdots, e^{\jmath\omega_{N_{\rm x}}}]^{T}\in\mathbb{C}^{N_{\rm x}\times 1}$ and ${\bf v}_{\rm z}=[e^{\jmath\omega_1}, e^{\jmath\omega_2}, \cdots, e^{\jmath\omega_{N_{\rm z}}}]^{T}\in\mathbb{C}^{N_{\rm z}\times 1}$ are referred to as the horizontal and vertical IRS beam training vectors, respectively. As such, ${\bf c}^{H}_{k} {\bf v}$ in \eqref{Eq:Signal} can be rewritten as
 \vspace{-2pt}
\begin{align}
{\bf c}^{H}_{k} {\bf v}&= \l({\bf u}^{H}(\tilde{\varphi}_{{\rm I},k}, N_{\rm x}) \otimes  {\bf u}^{H}(\tilde{\chi}_{{\rm I}}, N_{\rm z})\r) ({\bf v}_{\rm x}\otimes {\bf v}_{\rm z})\nn\\
&=\l({\bf u}^{H}(\varphi_{{\rm I},k}, N_{\rm x}){\bf v}_{\rm x}\r)\otimes \l(  {\bf u}^{H}(\chi_{{\rm I}}, N_{\rm z}){\bf v}_{\rm z}\r),\label{Eq:3D_decouple}
\end{align}
where  the horizontal and vertical beam training vectors are decoupled. For simplicity, we assume that the IRS vertical beamforming has been aligned as it does not depend on users' locations and thus focus on  designing the IRS horizontal  beam training for all the users. Specifically, given the fixed ${\bf v}_{\rm z}$ and using \eqref{Eq:3D_decouple}, the received signal at each user $k$ in \eqref{Eq:Signal} can be simplified as
\vspace{-5pt}
\begin{align}
y_k({\bf v}_{\rm x})&= \l({\bf u}^{H}(\varphi_{{\rm I},k}, N_{\rm x}){\bf v}_{\rm x}\r) \zeta_k  x +{n}_k,
~~\forall k\in\mathcal{K},\label{Eq:Signal2}
\end{align}
where $\zeta_k= {\bf u}^{H}(\chi_{{\rm I}}, N_{\rm z}){\bf v}_{\rm z}\eta_k$. 

\vspace{-5pt}
\section{Single-Beam Training}
Similar to \cite{xiao2016hierarchical}, given $N_{\rm x}$ IRS horizontal reflecting elements, we divide the entire spatial domain $[-1, 1]$ into $J\triangleq N_{\rm x}$ equal-size sectors for the horizontal beam training, represented by their central directions that  are given by $\alpha(j)= -1+\frac{2j  -1}{N_{\rm x}}, j \in\mathcal{J}\triangleq\{1, 2,\cdots, J\}$. 
As such, the single-beam training codebook  can be constructed as ${ \bf{\widetilde W}}=\{{\bf \widetilde w}(1), {\bf \widetilde w}(2), \cdots, {\bf \widetilde w}(J)\}$, where $\widetilde{\bf w}(j)\in\mathbb{C}^{N_{\rm x}\times 1}, j\in\mathcal{J}$ denotes the codeword that steers reflecting beam towards  direction $\alpha(j)$, which can be set as \cite{xiao2016hierarchical} ${\widetilde{\bf w}}(j)={\bf u}(\alpha(j), N_{\rm x})
$. Let ${\bf A}(\widetilde{\bf w}(j), \varphi)=|{\bf u}^{H}(\varphi, N_{\rm x}) \widetilde{\bf w}(j) |$ denote the beam gain of $\widetilde{\bf w}(j)$ along the spatial direction $\varphi\in[-1, 1]$.  It is well known that  the beam pattern of $\widetilde{\bf w}(j)$ (i.e., $\{{\bf A}(\widetilde{\bf w}(j),  \varphi)|~ \varphi\in[-1, 1]\})$  has a main-lobe with beam width $2/N_{\rm x}$ centered at the direction $\varphi=\alpha(j)$, where it achieves the maximum beam gain of ${\bf A}(\widetilde{\bf w}(j), \alpha(j))=N_{\rm x}$ \cite{xiao2016hierarchical}. Moreover, as $N_{\rm x}$ increases, the main-lobe becomes narrower and the side-lobe diminishes. 

Given the sampled directions, we denote by $I_k$ the optimal  IRS beam direction for each user $k$, which is given by $I_k=\arg\min_{j \in\mathcal{J}}\l|\varphi_{{\rm I},k}-\alpha(j)\r|, \forall k\in\mathcal{K}$.  With the codebook $\widetilde{\bf{W}}$, a straightforward IRS beam-training method is as follows: The AP consecutively sends multiple training symbols while the IRS changes its reflecting  direction in $\{{\widetilde{\bf w}}(j), j\in\mathcal{J}\}$ sequentially over different training symbols;  then each user finds its best  beam direction that achieves the maximum received signal power/SNR, which is given by $\widehat{I}^{(\rm ex)}_k=\arg\max_{j \in\mathcal{J}}|y_k(\widetilde{\bf w}(j))|^2, \forall k\in\mathcal{K}$.
However, such an exhaustive-search based single-beam training requires at least  $T^{(\rm ex)}_{\rm t}=N_{\rm x}$ training symbols, which can be  practically prohibitive  due to the massive number of IRS reflecting elements, thus incurring large training overhead/delay for establishing high-SNR links. As such, this training method is not suitable for delay-sensitive and/or short-packet transmissions.

\vspace{-10pt}
\section{Multi-Beam Training}
To reduce the training time of conventional single-beam training, we propose  a new multi-beam training method for IRS-assisted multiuser communications in this section. 

\begin{figure*}[t]
\begin{center}
\includegraphics[height=5.5cm]{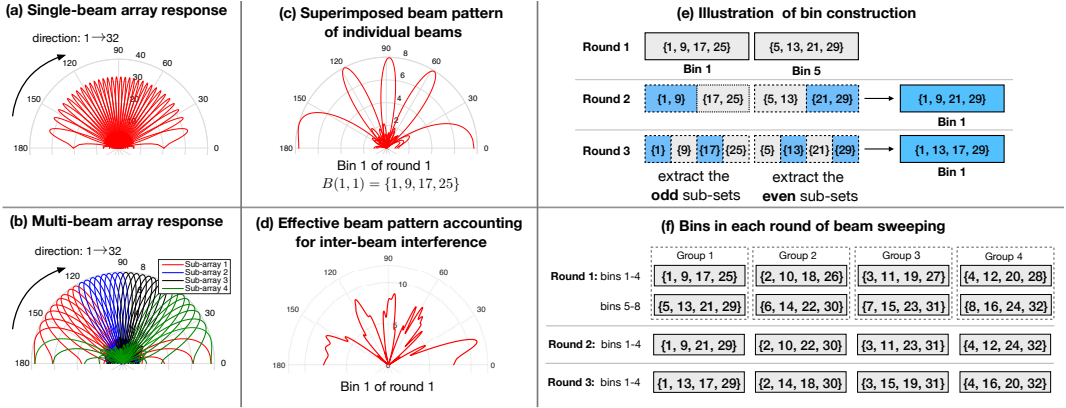}
\caption{Illustration of the proposed IRS multi-beam training with $N_{\rm x}=32$, $M=4$, $L=8$, and $d_{\rm I}=\lambda/2$.}
\label{Fig:BeamExample}
\end{center}
\vspace{-7pt}
\end{figure*}

First, we divide  the  (horizontal) IRS reflecting elements into $M$
 sub-arrays, each  consisting of  $L\triangleq N_{\rm x}/M$ (assumed to be an integer) adjacent reflecting elements.
 For each sub-array $m\in\mathcal{M}\triangleq\{1, 2, \cdots, M\}$, we equip it with an individual codebook, ${\bf W}_m$, which comprises $J=N_{\rm x}$ codewords that cover the same sampled directions as the single-beam codebook, i.e., $\{\alpha(j), j\in\mathcal{J}\}$. As such,  we have  ${\bf W}_m=\{{\bf w}_m(1), {\bf w}_m(2), \cdots, {\bf w}_m(J)\}, \forall m\in\mathcal{M},$ where ${\bf w}_m(j)\in\mathbb{C}^{L\times 1}$, $j \in\mathcal{J}$ represents the codeword of sub-array $m$ that steers reflecting beam in direction $\alpha(j)$ using $L$ reflecting elements only. 
Based on the single-beam codeword ${\bf \widetilde w}(j)$, we construct ${\bf w}_m(j)$ as
\vspace{-4pt}
 \begin{align}
 {\bf w}_m(j)\triangleq[{ \widetilde{\bf w}(j)}]_{(m-1)L+1: mL}=e^{\jmath (m-1)L \alpha(j)} {\bf u}(\alpha(j), L),\nn
 \vspace{-3pt}
 \end{align}
 such that when all the $M$ sub-arrays steer reflecting beams in the same direction $\alpha(j)$, the composite multi-beam codeword $[{\bf w}^{T}_1(j), {\bf w}^{T}_2(j), \cdots, {\bf w}^{T}_M(j)]^{T}$ is equivalent to the single-beam counterpart $\widetilde{\bf w}(j)$.
Compared to the full-array codeword $\widetilde{\bf w}(j)$, each  sub-array codeword ${\bf \widetilde w}_m(j)$ has a wider beam width (i.e., $2M/N_{\rm x}$ versus $2/N_{\rm x}$) as well as a smaller beam gain (i.e., $N_{\rm x}/M$ versus $N_{\rm x}$), as illustrated in Figs.~\ref{Fig:BeamExample}(a) and \ref{Fig:BeamExample}(b), respectively.

Next, for the proposed multi-beam training, let the IRS steer sub-array beams towards multiple different   directions simultaneously which generally change over training symbol durations according to the multi-beam  codebook $\{{\bf W}_m, m\in\mathcal{M}\}$.
By properly designing the beam directions of IRS sub-arrays over different training symbols, each user's optimal beam direction can be found via simple received signal  power/SNR comparisons over time  with a high probability.  
For ease of exposition, we consider a typical case where $M=2^{R}$ with $R\in\mathbb{Z}$,  $L$ is an even number, and $N_{\rm x}= ML$. Our proposed  fast beam-training method consists of two phases, namely, IRS beam sweeping and IRS beam identification, which are elaborated as follows.  

\subsubsection{\textbf{IRS Beam Sweeping}}

This phase consists of $1+\log_2 M$ rounds of beam sweeping, where in each round, the AP sends multiple training symbols that are reflected by different sets of IRS sub-array reflecting beam directions. For each round $r\in\{1, 2, \cdots, \log_2 M+1\}$, we denote by $B(r, b)$ the bin that collects the sub-array beam directions (arranged in an ascending order) during the $b$-th training symbol.  For any beam-direction set $\mathcal{A}\subseteq\mathcal{J}$, we define its \emph{intra-set distance} as $d_{\rm s}(\mathcal{A})=\min_{p,q\in \mathcal{A}; p\neq q} |p-q|$. As such, a larger intra-set distance indicates that the beams indexed by $\mathcal{A}$ are farther  separated in the spatial domain (see Fig.~\ref{Fig:BeamExample}(b)). 
For $r=1$, we map the $N_{\rm x}$ directions into $L$ bins, each comprising $M$ directions. To separate the beam directions in each bin as far as possible for minimizing the inter-beam interference, we set the bins as $B(1, b)=\{b, b+L, \cdots, b+(M-1)L\}, \forall b\in\{1, 2, \cdots, L\}$. It can be shown that such a direction-bin mapping maximizes the minimum intra-bin distance among all the bins with $d_{\rm s}(B(1, b))=L, \forall b\in\{1, 2, \cdots, L\}$. As illustrated in Figs.~\ref{Fig:BeamExample}(c) and \ref{Fig:BeamExample}(d), for the case with $N_{\rm x}\!=\!32$ and $M\!=\!4$,  the individual beam patterns of sub-array beams in $B(1, 1)=\{1, 9, 17, 25\}$ (corresponding to the codeword ${\bf v}_{\rm x}=[{\bf w}^{T}_1(1),{\bf w}^{T}_2(9), {\bf w}^{T}_3(17), {\bf w}^{T}_4(25)]^T$) are well separated in the spatial domain (see Fig.~\ref{Fig:BeamExample}(c)), such that the effective beam pattern even after  accounting for the inter-beam interference still features strong beam directionality to the $4$ directions (see Fig.~\ref{Fig:BeamExample}(d)).

In the subsequent $(\log_2 M)$-rounds of  beam sweeping, we exploit different combinations of IRS beam directions to help each user identify its best beam direction in the next beam identification phase.   Specifically, in round $2$, for each initial bin $B(1,b)$ with $b\in\{1,2, \cdots, L\}$, we partition its $M$ directions into $2$ equal-size sub-sets as $[B(1,b)]_{1:M/2}$ and $[B(1,b)]_{(M/2)+1:M}$, each consisting of adjacent $M/2$ directions.  To determine which sub-set contains the best beam direction, instead of individually beam-searching  the $2$ sub-sets and comparing their beam power, we propose to test only  one of them (with half number of sub-arrays) for reducing the training time, by exploiting  the fact that the searched  sub-set is likely to contain the best beam direction if it yields a large received power above a certain threshold (as specified later in the next subsection), and vice versa. Moreover, to further reduce the training time, we make full use of the $M$ sub-arrays to allow simultaneous search of two sub-sets from two different initial bins, which, however, introduces the interference  between different beam sub-sets. To address this issue, we first pair the initial bins into $L/2$ groups as $G(\ell)=\{B(1, \ell), B(1, \ell+L/2)\}, \forall \ell\in[1, 2, \cdots, L/2]$, such that the beam directions in the two bins for all groups are separated as far as possible with the identical (maximum) intra-group distance  given by $d_{\rm s}(G(\ell))=L/2, \forall \ell\in[1, 2, \cdots, L/2]$. Then, for each group $\ell$, we extract the odd sub-set of $B(1, \ell)$ and the even sub-set of $B(1, \ell+L/2)$ to construct a new bin (as illustrated in Fig.~\ref{Fig:BeamExample}(e)). As such,  $L/2$ bins are constructed  for $r=2$, 
which are set as $B(2, b)=\{[B(1, b)]_{1:M/2}, [B(1, b+L/2)]_{(M/2)+1: M}\}, \forall b\in[1, 2, \cdots, L/2].$ 
It can be verified that by using the above bin \emph{grouping-and-extracting}, the second-round of beam sweeping  achieves the same max-min intra-bin distance as the first round (i.e., both are $L$).
Similarly, for round $3$, as illustrated in Fig.~\ref{Fig:BeamExample}(e), we further partition each of the sub-sets in round $2$ into $2$ equal-size smaller sub-sets and select one of them for beam searching; the new bins are constructed by following a similar procedure in round $2$.  To summarize, for $r\in[2, 3, \cdots, \log_2 M+1]$, each round of beam sweeping consists of $L/2$ bins, which are set as
\vspace{-8pt}
\begin{align}
&B(r, b)=\{[B(1, b)]_{1:u(r)}, [B(1, b+L/2)]_{u(r)+1: 2 u(r)},\nn\\
\vspace{-2pt}
&\qquad\qquad~[B(1, b)]_{2u(r)\!+\!1: 3 u(r)}, [B(1, b+L/2)]_{3u(r)\!+\!1: 4 u(r)},\nn \\
&\qquad\qquad~~~\cdots,[B(1, b)]_{M\!-\!2u(r)\!+\!1: M-u(r)}, \nn\\
&\qquad\qquad[B(1, b+L/2)]_{M\!-\!u(r)+1: M} \}, \forall b\in[1, 2, \cdots, L/2],\nn
\vspace{-3pt}
\end{align}
where   $u(r)=M/(2^{r-1})$. Moreover, for each round of beam sweeping, the identical (maximum) intra-bin distance for different bins can be numerically deduced as
\vspace{-5pt}
\begin{align}
d_{\rm s}(B(r, b))=\begin{cases}
&\!\!\!\!\!L, ~~~~\mbox{if $r \in\{1, 2\}$},\nn\\
&\!\!\!\!\!L/2, ~\mbox{if $r \in\{3, 4, \cdots, \log_2 M+1\}$}.
\end{cases} 
\vspace{-5pt}
\end{align} 
 For illustration, we provide in Fig.~\ref{Fig:BeamExample}(f) an example of all the designed bins in the beam-sweeping phase for the case with  $N_{\rm x}=32$ and $M=4$. 

Based on the above beam-sweeping design, the total number of training symbols  of  our proposed multi-beam training method is given by
\vspace{-7pt}
\begin{equation}
T^{(\rm fa)}_{\rm t}=L+\frac{L(\log_2 M)}{2}=\frac{N_{\rm x}}{M}\l(1+\frac{\log_2 M}{2}\r),\label{Eq:TrainOver}
\vspace{-5pt}
\end{equation} 
which monotonically decreases with an increasing $M$ (i.e., IRS reflecting elements are divided into more sub-arrays), and  is smaller than that of the single-beam training 
 with  $T^{\rm (ex)}_{\rm t}=N_{\rm x}$ for $M>1$.

\subsubsection{\textbf{IRS Beam Identification}} \label{Sec:Deter}
After the  beam-sweeping phase, each user can identify its best IRS reflecting beam direction \emph{independently} based on their own received powers/SNRs in the first phase. Consider an arbitrary user $k$. Let $P_k(r,b)$ denote its received  power from  the $b$-th bin of the $r$-th round of beam sweeping and $\widehat{I}_{k}(r)$ represent the set of candidate directions for its best beam direction  after the $r$-th round of beam sweeping.  For $r=1$, $\widehat{I}_{k}(r)$ is set as the best bin that has  the largest  received power, i.e.,  $\widehat{I}_k(1)=B(1, b^*_k)$ with $b_k^*=\arg\max_{b\in\{1, 2, \cdots, L\}} P_k(1,b)$. While for each of the subsequent rounds $r\in\{2, 3, \cdots, \log_2{M}+1\}$, the user  only needs to inspect one bin that has common directions with $\widehat{I}_k(1)$ as only it may potentially   contain  the best beam direction, which is denoted by $B(r, b_k(r))$ with $b_k(r)=\{b\in\{1,2,\cdots, L/2\}| B(r, b)\cap \widehat{I}_k(1)\neq \varnothing\}$. For this bin, as the expected received power from the corresponding multiple beams that cover/do not cover the best direction is approximately (ignoring the receiver noise and any inter-beam interference) $P_k(1,b_k^*)$ and $0$, respectively, we set the \emph{binary-decision} threshold on the received power as $P^{(\rm th)}_k\triangleq  (P_k(1,b_k^*)+0)/2=P_k(1,b_k^*)/2$ for determining whether $B(r, b_k(r))$ contains the best beam direction or not.
As such, for each $r\in\{2, 3, \cdots, \log_2{M}+1\}$, combining the binary decision with $\widehat{I}_k(r-1)$, the new candidate directions for round $r$ are determined as follows.
\vspace{-4pt}
\begin{equation}
\widehat{I}_k(r)\!\!=\!\!
\begin{cases}
&\!\!\!\!\widehat{I}_k(r-1)\!\cap\! B(r, b_k(r)), ~\mbox{if  $P_k(r,b)\!\ge\! P^{(\rm th)}_k$},\nn\\
&\!\!\!\!\widehat{I}_k(r-1)\!\setminus\! B(r, b_k(r)), ~\mbox{if  $P_k(r,b)\!<\! P^{(\rm th)}_k$},\nn\\
\end{cases} \forall k\in\mathcal{K}.
\vspace{-8pt}
\end{equation}
It can be verified that  the size of $\widehat{I}_k(r)$ is logarithmically decreasing as $|\widehat{I}_k(r)|=M/(2^{r-1}), \forall r\in\{1, 2, \cdots, \log_2{M}+1\}$.
An illustrative example is provided as follows to demonstrate the detailed procedures for identifying a unique beam direction  for an arbitrary user. 
\begin{example}
\vspace{-4pt}
\emph{Consider the case with $N_{\rm x}=32$ and $M=4$; the designed bins are shown in Fig.~\ref{Fig:BeamExample}(f). For $r=1$, assuming  that user $k$ receives the largest power from bin $B(1,5)$, we set $\widehat{I}_{k}(1)=B(1,5)=\{5, 13, 21, 29\}$. 
As such, for $r=2$, the user only needs to examine bin $B(2,1)$, since only it has common directions with bin $B(1,5)$. Supposing $P_k(2,1)<P^{(\rm th)}_k$, we decide that $B(2,1)$ does not contain the best beam direction of user $k$ and thus  obtain $\widehat{I}_{k}(2)=\widehat{I}_{k}(1)\setminus B(2,1)=\{5, 13\}$. 
Last, for $r=3$, user $k$ only needs to inspect bin $B(3,1)$. Assuming $P_{k}(3,1)> P^{(\rm th)}_k$, we finally identify the best beam direction for user $k$ as $\widehat{I}_{k}^{(\rm fa)}=\widehat{I}_{k}(3)=\widehat{I}_{k}(2)\cap B(3,1)= \{13\}$. }
\vspace{-3pt}
\end{example}

Note that for each user $k$, the identified best beam direction, $\widehat{I}_{k}^{(\rm fa)}$, may not be the actual optimal beam direction,  $I_{k}$, or that obtained by the single-beam training, $\widehat{I}_{k}^{(\rm ex)}$, due to the receiver noise, interference due to channel non-LoS (NLoS) components,  and inter-beam interference in practice. {Although increasing $M$ can help reduce the training time of the proposed multi-beam training method (see \eqref{Eq:TrainOver}), it will decrease the sub-array beam gains as well as cause more severe inter-beam interference, thus resulting in degraded beam identification accuracy, as will be shown in the next section by simulations. Hence, there exists a fundamental trade-off between training time and resultant passive beamforming performance in the proposed multi-beam training method by adjusting $M$.}

\vspace{-10pt}
\section{Simulation Results}\label{Sec:Simu}
\vspace{-4pt}

This section provides simulation results to numerically validate our proposed design.
We consider a mmWave system operating at a carrier frequency of $30$ GHz. For simplicity, we consider an IRS array   placed horizontally and centered at $(0, 0, 0)$ meter (m), which is composed of $N_{\rm x}=160$ reflecting elements
with $d_{\rm I}=\lambda/4$. There are $K=5$ users randomly distributed on a semi-circle around the IRS with distance of $2$~m. The AP centered at $(0,16,0)$ m is equipped with $N_{\rm A}=64$ antennas with half-wavelength antenna spacing. 
For the large-scale path loss, the reference channel power gain at a distance of $1$ m is set as $\xi_0=-62$ dB, and the path loss exponents of the AP-IRS and IRS-user links are set as $\gamma_{\rm AI}=2.3$ and $\gamma_{\rm IU}=2$, respectively. The small-scale fading is modeled by the Rician fading, with the AP-IRS and IRS-user Rician factors set as $\kappa_{\rm AI}=5$ dB and $\kappa_{\rm IU}=10$ dB, respectively. 
We define the average SNR of the IRS-assisted mmWave system as 
\vspace{-6pt}
\begin{equation}
{\rm SNR}=\frac{P_{\rm A}(\xi_0 D_{\rm AI}^{-\gamma_{\rm AI}}) (\xi_0 D_{\rm IU}^{-\gamma_{\rm IU}})N^2_{\rm x} N_{\rm A}}{\sigma^2},
\vspace{-5pt}
\end{equation}
where $D_{\rm AI}$ and $D_{\rm IU}$ denote respectively the AP-IRS and IRS-user distances, and the noise power is set as $\sigma^2=-109$ dBm.  To characterize the beam identification accuracy, we define $P_{\rm suc}=\frac{\sum_{k=1}^K\mathbb{I}(\widehat{I}_k=I_k)}{K}$ as its success rate,
 where $\mathbb{I}(\cdot)$ stands for the indicator function. Moreover, to show the passive beamforming gain for data transmission in the  identified single-beam direction, $\widehat{I}_{k}$, we define the average achievable rate of all users  as $\bar{R}=\frac{\sum_{k=1}^K R_k}{K}$, where
$R_k=\log_2 \l(1+\frac{P_{\rm A} | \l({\bf u}^{H}(\varphi_{{\rm I},k}, N_{\rm x}){\bf v}_{{\rm x},k}\r) \zeta_k  |^2}{\Gamma \sigma^2}\r)$, ${\bf v}_{{\rm x},k}={\bf u}(\alpha(\widehat{I}_{k}), N_{\rm x})$, and $\Gamma=9$ dB
 denotes the SNR gap due to the practical modulation and coding. Note that we ignore the loss of achievable rate due to training overhead for ease of comparison.  The simulation results are averaged over $1500$ Rician fading channel realizations.

\begin{figure}[t]
\centering
\subfigure[Success beam identification rate versus SNR for different numbers of IRS sub-arrays.]{\label{FigAccu_arraySize_cont3}
\includegraphics[height=3.6cm]{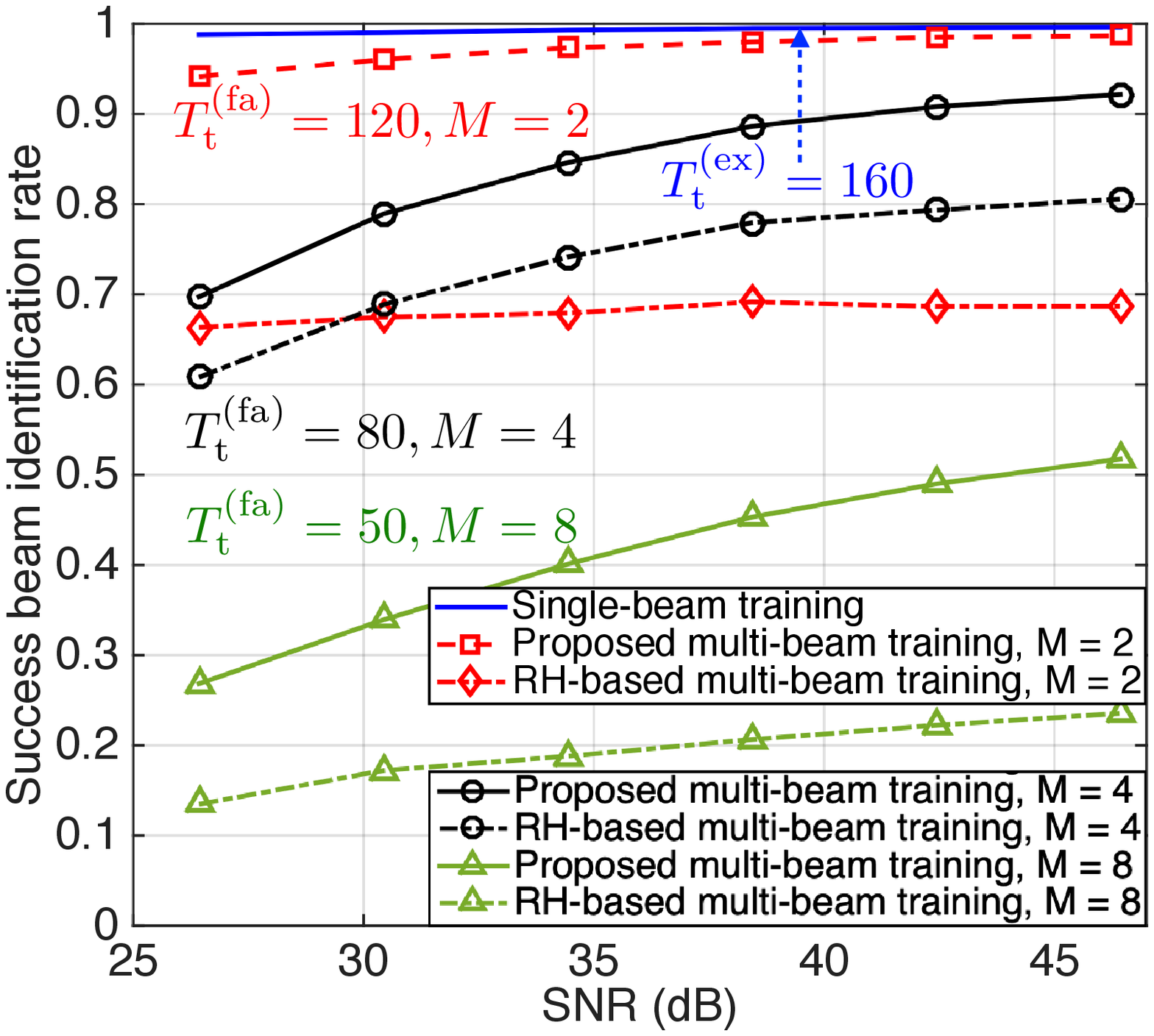}}
\hspace{5pt}
\subfigure[Average achievable  rate versus SNR for different numbers of IRS sub-arrays.]{\label{FigRate_arraySize_cont2}
\includegraphics[height=3.6cm]{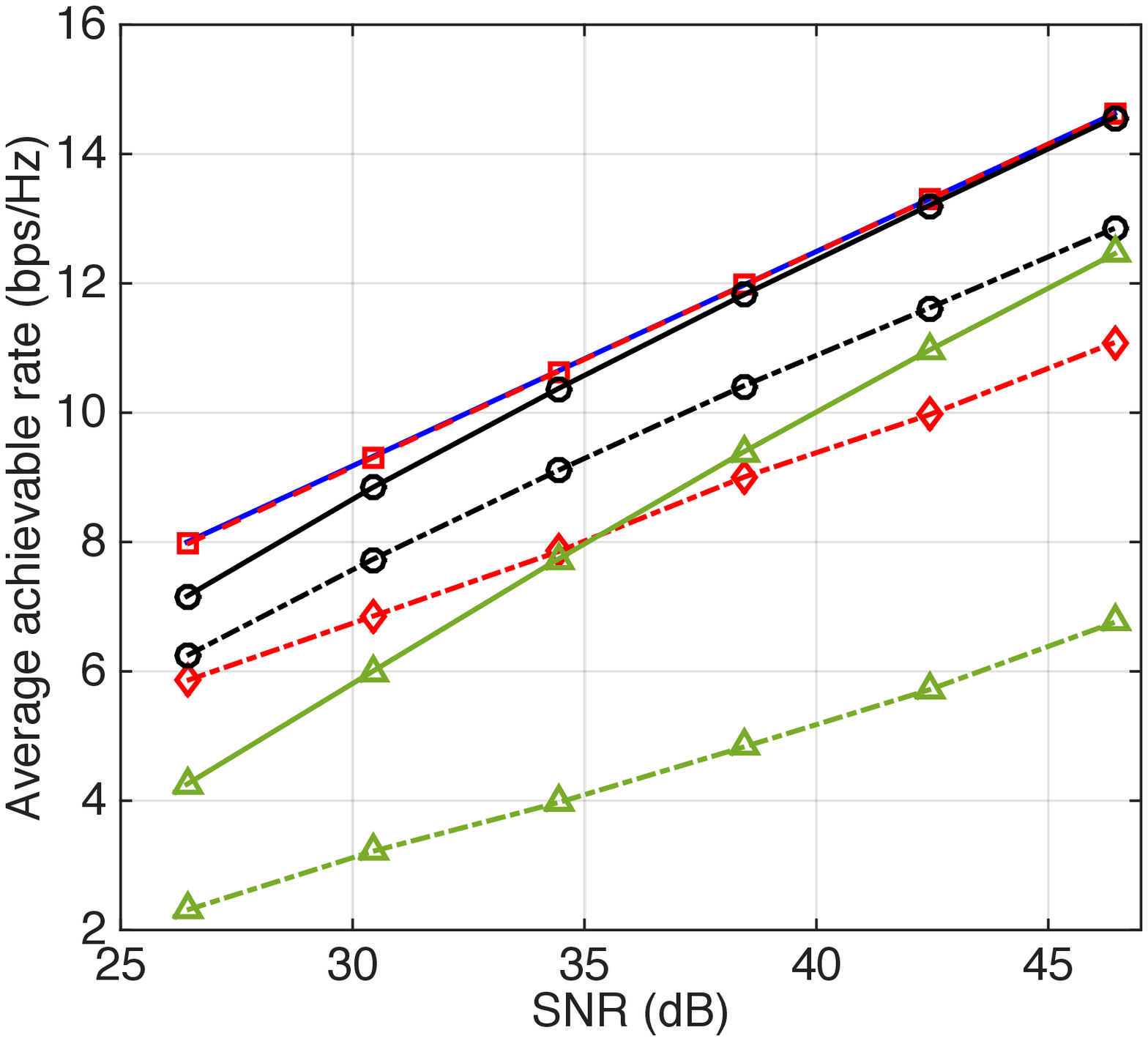}}
\label{Fig:Rate}
\caption{Performance comparison of the proposed multi-beam training with the conventional single-beam training and  RH based multi-beam training}.
\vspace{-5pt}
\end{figure} 

Figs.~\ref{FigAccu_arraySize_cont3} and \ref{FigRate_arraySize_cont2}
show the effects of the number of IRS sub-arrays ($M$) and SNR on the training overhead, success beam identification rate, and  average achievable rate. The proposed multi-beam training is compared with the conventional single-beam training as well as the RH based multi-beam training proposed in \cite{hassanieh2018fast} that conducts the max-min intra-bin distance direction-bin mapping followed by multiple rounds of RH for beam sweeping, and applies a voting mechanism for beam identification. For fair comparison, we enforce the same training time for the RH based multi-beam training as our proposed one by limiting its number of direction-bin mappings, so as to compare their beam identification accuracy and passive beamforming gain with the same training overhead.
 Several interesting observations are made as follows. {First,  
 with a small number of sub-arrays (i.e., $M=2$), the proposed multi-beam training not only reduces $25\%$ of the training overhead of the single-beam training (i.e., $120$ versus $160$), but also achieves a very close success beam identification rate and average achievable rate. 
 Second, by slightly increasing the number of sub-arrays to $M=4$, the proposed multi-beam training achieves $50\%$ 
 training time reduction with respect to the single-beam training (i.e., $80$ versus $160$), while 
 it still attains a high success beam identification rate at high SNR  (e.g., $P_{\rm suc}\approx92\%$ for SNR = $46.4$ dB) as well as close rate performance to the single-beam training (see Fig.~\ref{FigRate_arraySize_cont2}). 
 However, the proposed multi-beam training with $M=8$ is observed to suffer a substantial loss in the  beam identification accuracy even in the high-SNR regime (see Fig.~\ref{FigAccu_arraySize_cont3}) and thus degraded passive beamforming gain (see Fig.~\ref{FigRate_arraySize_cont2}), due to more severe inter-beam interference.} Moreover, it is observed that  the proposed multi-beam training significantly outperforms the RH based benchmark for all values of $M\in\{2,4,8\}$ in terms of both the beam identification accuracy and passive beamforming gain, owing to its more efficient beam sweeping and identification designs. In particular, the beam identification accuracy of the RH based multi-beam training with $M\!=\!2$ is almost invariant with the increase of SNR, since its RH round in beam training only randomly covers half of the total beam directions,
 thus inevitably missing some directions and greatly limiting the voting-based beam identification performance.

\vspace{-10pt}
\section{Conclusions}
\vspace{-3pt}
In this letter, we proposed a fast IRS reflect beam-training method for an IRS-assisted multiuser communication system. 
It was shown that by dividing IRS elements into multiple sub-arrays and properly designing sub-array beam directions over different training symbols with users' independent beam identification based on received power/SNR comparisons,  
our proposed multi-beam training can significantly reduce the training overhead of conventional single-beam training, yet achieving comparable passive beamforming performance for data transmission.   
Moreover, it is worth noting that the proposed multi-beam training method is general and can also be applied to IRS's vertical beam training as well as  AP's transmit beam training  to multiple users without IRS or with fixed IRS (horizontal)  reflection.

\end{document}